# Optical Trapping Based on Microring Resonators with Transverse Slot Structure

ZHENG LI,[1] YI CHENG,[2] JIN LIU,[2] GUANJU PENG[1,*]

[1]School of Precision Instruments and Optoelectronic Engineering, Tianjin University, Tianjin 300387, China
[2]Shanghai Institute of Satellite Engineering, Shanghai 201109, China
*Corresponding author: guanjup@tju.edu.cn

**Over the past few decades, optical manipulation has emerged as a highly successful tool in various fields, such as biology, micro/nanorobotics, and physics. Among the different techniques, the transverse slot optical waveguide has shown remarkable potential in enhancing the field and significantly improving optical trapping capabilities. Additionally, microring resonators have demonstrated the ability to enhance the field at specific resonance wavelengths, enabling the manipulation and capture of particles. In this study, we investigated the impact of the structure on nanoparticle capture by introducing a 50 nm transverse slot in a 5 μm microring resonator. Through the integration of a transverse slot in the microring resonator, we observed a substantial increase in the maximum bound optical power for a nanosphere with a refractive index of 1.6 and a diameter of 50 nm, reaching 3988.8 pN/W. This value is 2292 times higher than the maximum optical force in a straight waveguide and 2.266 times higher than the maximum optical force in a microring resonator. The proposed structure significantly enhances the optical trapping capabilities for nanoscale particles, thus paving the way for the development of advanced micro/nanomanipulation techniques. © 2023 Optica Publishing Group**

## 1. INTRODUCTION

The study of the microcosmic world in modern life sciences has extended to the nanoscale, attracting considerable attention due to the non-invasive and non-contact nature of optical trapping technology [1-4]. The trapping force in optical trapping is directly influenced by the cube of the particle radius and the gradient of the optical field. However, traditional optical traps [5-7], constrained by the diffraction of free-space laser beams, encounter difficulties in precisely manipulating and controlling nanoparticles smaller than 100 nm. In contrast, evanescent field optics presents a promising solution by overcoming these limitations [8]. Among the excellent platforms for investigating evanescent field optical confinement, silicon-based waveguides stand out, enabling the exploration of the fundamental characteristics of evanescent field optical forces [9-14] while offering exceptional versatility.

Photonics-based evanescent-field optical confinement devices can be broadly classified into three categories: waveguide-based [15-23], resonator-based [24-27], and surface plasmon-enhanced [28-31]. However, all three types share a common foundation in waveguide structures. By leveraging well-designed structures, these evanescent field optical confinement devices can achieve field enhancement in specific regions, effectively reducing the need for high incident light power while overcoming the limitations imposed by the diffraction limit. Moreover, these devices exhibit compact footprints, seamless integration capabilities, and compatibility with photonic chip platforms. As a result, their research is closely intertwined with integrated optics, underscoring their significant research significance.

Ng successfully manipulated gold spheres with a radius of 10 nm utilizing evanescent fields generated around the waveguide, achieving an impressive maximum transport speed of 4 μm/s [8]. Similarly, Almeida introduced a slotted waveguide design featuring a 50 nm-wide slot positioned in the middle of the straight waveguide [13]. This innovative structure enables the confinement of the field within a narrow region of low refractive index, resulting in an intensity approximately 20 times greater than that found in a conventional straight waveguide. Yang conducted successful trapping experiments on nanoparticles with diameters of 300 nm and 600 nm using the evanescent field generated around a straight waveguide with light transmission [14]. In a separate study, Yang introduced a longitudinal slot measuring 60 nm in the straight waveguide. This design compressed the light within the narrow longitudinal waveguide, enabling the capture and transport of 75 nm dielectric nanoparticles and λ-DNA molecules [15].

In another investigation, Driscoll incorporated a lateral slot of 50 nm in a straight waveguide, resulting in the confinement of 97% of the transverse field component within the slot. This arrangement led to the achievement of a significant trapping force [16]. Hoshina proposed a nanophotonic tweezer that leverages optical nonlinear effects in waveguides to enable super-resolution trapping [17]. Shi introduced an array of optical potential wells capable of highly selective, large-scale trapping, and nanoscale sorting [22]. Additionally, Fang utilized subwavelength longitudinal slot waveguides for the separation of chiral nanoparticles [23]. The particle of less than 100 nm has been trapped by off-plane solutions. Some researchers had achieved nanoscale optical trapping by a silicon-based dielectric nanobowtie dimer and photonic nanojets [32,33]. Meanwhile, Pang achieved the optical trapping of a single molecule that has a radius of 3.4 nm, using a double-nanohole in an Au film [34].

The resonance of a microring resonator leads to a significant enhancement of the light confined within the ring, resulting in an increased optical field gradient along the ring. This enhanced gradient facilitates the capture of particles. Yang and Erickson were the first to report the utilization of microring resonators positioned adjacent to straight waveguides for the confinement of microscale particles [35]. Geng employed finite-difference time-domain simulations to investigate the optical forces and manipulation capabilities of coupled double waveguide ring resonators. This research laid the groundwork for exploring optically driven reconfigurable optical functional components [36]. Kotsifaki proposed a metamaterial-assisted optical tweezer that leverages the Fano resonance effect for the capture of individual nanoparticles [37]. Furthermore, Zhou achieved the coupling of single atoms to a nanophotonic echo chamber resonator via waveguides [38].

Zhou demonstrated the resonance thermal sensitivity of the slotted silicon microring resonators [39]. Fang's numerical investigations shown that the slot waveguide can offer high efficiency and feasibility of separating chiral nanoparticles [40]. Moreover, the multi-trapping was widely researched by dielectric metasurfaces [41] and plasmonic nanostructures [42]. Theoretically, the multi-trapping could be achieved through placing several slots within the ring resonator. Thus, the slotted microring resonators can assess multiple trapping, but this is beyond the scope of this paper.

The existing research either studied the optical trapping of slotted straight waveguides, the optical trapping of microring resonators, or studied the resonance thermal sensitivity of the slotted microring resonators. Both the microring resonators and the slot have a strong trapping effect on nanoparticles, and the combination of the two to achieve optical trapping may achieve some excited results. As far as we know, the slotted microring resonators for optical trapping has not been reported.

Considering the significant confinement effects offered by slots and microring resonators on light, this study introduces a lateral slot with a width of 50 nm in an all-pass microring resonator. The investigation utilizes theoretical analysis and finite element analysis software to explore the collective confinement effects of the microring resonator and slot on light and to examine the trapping capability of this structure on nanoparticles.

## 2. METHODS

### 2.1 The principle of bound light through a slot waveguide

The boundary conditions at a discontinuous interface can be derived from Maxwell equations [43],

$$\vec{D}_{2n} - \vec{D}_{1n} = \sigma_f, \quad (1)$$

where, $D$ is the electric displacement vector, and $\sigma_f$ is the charge surface density of free charges. Due to the absence of free charges on both sides of the interface between the media, the vector of electric displacement remains continuous across the media on either side.

$$\vec{D}_{2n} = \vec{D}_{1n} = \varepsilon_0 \varepsilon_{r1} \vec{E}_{1n} = \varepsilon_0 \varepsilon_{r2} \vec{E}_{2n}, \quad (2)$$

where, $\varepsilon$ is the permittivity, and $E$ is the electric field. Since $\varepsilon_0\varepsilon_r$ is directly proportional to $n^2$,

$$\varepsilon_0 \varepsilon_{r1} \vec{E}_{1n} = \varepsilon_0 \varepsilon_{r2} \vec{E}_{2n} \Rightarrow n_1^2 \vec{E}_{1n} = n_2^2 \vec{E}_{2n}, \quad (3)$$

where, $n_1$ and $n_2$ are the refractive indices on both sides of the interface. In the case of the interface between a waveguide and a slot, the media on either side of the interface are the optical waveguide and the cladding.

Here, the waveguide material in this article is silicon, with a refractive index of approximately 3.5, while the cladding is water, with a refractive index of about 1.33, in 10 °C [44]. A significant refractive index difference arises across the interface, as illustrated in Fig. 1(a). The $x$ direction is defined as the vertical direction (height) and the $y$ direction is the horizontal direction (width). Light enters from the end face of the waveguide on the left, enters the waveguide on the right through a slot with a low refractive index, and finally exits, as shown in Fig. 1(a).

According to the theory of Almeida[13], the electric field strength ratio on both sides of the interface can be calculated by

$$\frac{\vec{E}_{2n}}{\vec{E}_{1n}} = \frac{n_1^2}{n_2^2} = \frac{3.5^2}{1.33^2} \approx 6.93. \quad (4)$$

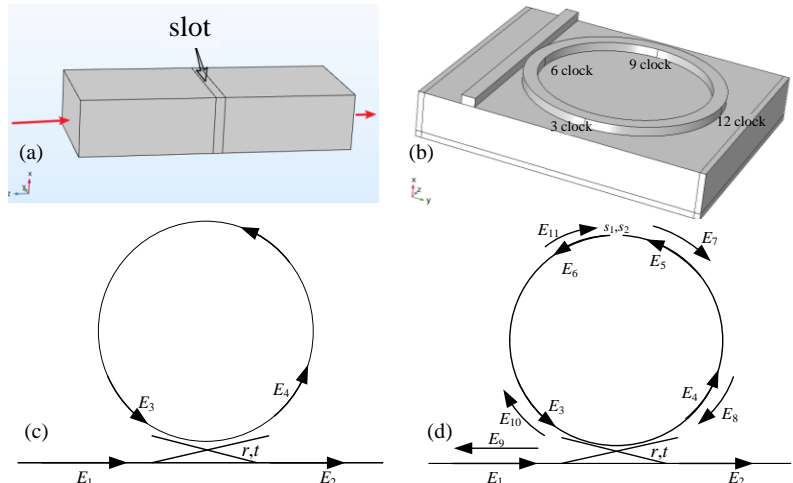

Fig.1. Schematic diagram of waveguide and microring. (a) waveguide; (b) microring resonator; microring resonator (c) and with slot (d).

The results showed that the light intensity in the slot was almost 6.9 times that in the optical waveguide. It can be assumed that most of the light is confined within the slot.

## 2.2 Microring resonator resonance theory

Here, we take the all-pass microring resonator as the research object, as shown in Fig. 1(b). the material of the ring and the straight waveguide is silicon, the substrate is silicon dioxide, and the cladding is a water layer.

In order to describe the various positions of the microring conveniently, we regard the microring resonator as a clock, and the coupling point close to the incident waveguide is defined as 6 clock. We divide the microring into 4 parts on average, and mark the four positions in the ring clockwise as 6 clock, 9 clock, 12 clock, 3 clock by clockwise. The resonance equation of light in the microring resonator is,

$$2\pi n_{eff} R = m\lambda \quad (m = 1, 2, 3...), \quad (5)$$

where $n_{eff}$ is the effective refractive index of the guided mode in the microring resonator, $R$ is the resonance radius, $m$ is the resonance order, and $\lambda$ is the wavelength of the resonant light. According to the literature [45], the scheme of the device shown in Fig. 1(c) is analyzed, and the ratio of the amplitude of the transmitted light to incident light can be calculated as,

$$\frac{E_2}{E_1} = \exp[i(\pi + \phi)] \frac{\tau - r\exp(-i\phi)}{1 - r\tau \exp(i\phi)}, \quad (6)$$

where $r$ is the self-coupling coefficient, $\tau$ is the amplitude transfer factor, $\phi = kL$ is the round-trip phase shift, $k = 2\pi n/\lambda$, $n$ is the effective refractive index of the guided mode inside the waveguide, $\lambda$ is the wavelength of the incident light, and $L$ is the waveguide ring perimeter.

The transmission can be calculated by solving for the square of ratio of the amplitude of the transmitted light to the amplitude of the incident light,

$$T = \left|\frac{E_2}{E_1}\right|^2 = \frac{\tau^2 - 2r\tau\cos\phi + r^2}{1 - 2r\tau\cos\phi + r^2\tau^2}. \quad (7)$$

When $r = \tau$, the transmission drops to 0. In this case, the internal loss is equal to the coupling loss, which is called the critical coupling. When $r > \tau$, the resonator is considered undercoupled, and when $r < \tau$, the resonator is considered overcoupled. The phase change between the transmitted light and the incident light can be calculated by Eq. (6),

$$\Phi = \arg\left(\frac{E_2}{E_1}\right) = \pi + \phi + \operatorname{atan}\left(\frac{r\sin\phi}{\tau - r\cos\phi}\right) + \operatorname{atan}\left(\frac{r\tau\sin\phi}{1 - r\tau\cos\phi}\right). \quad (8)$$

The free spectral range (FSR) in function of wavelength equals [46]:

$$FSR = \frac{\lambda^2}{n_{eff} L}, \quad (9)$$

where $L$ the circumference of microring resonators. The quality factor ($Q^2$) can be expressed as [46]:

$$Q^2 = \frac{\lambda}{\text{FWHM}}, \quad (10)$$

where, FWHM is the full width at half maximum of the resonant peak in microring resonators.

## 2.3 Theoretical calculation of microring resonator with slot at 12 clock

Consider the effect of designing a 50 nm slot at the 12 clock of the all-pass microring resonator on the microring resonator, as shown in Fig. 1(d). The slot size and nanoparticle size are not necessarily equal, when the slot size is 50 nm, the nanoparticle size can be less than 50 nm, or more than 50 nm, e.g., 30nm, 45 nm, 60 nm, and vice versa. Literature [46] shows that when the silicon microring resonators with 1.6 μm radius, the loss is 3 dB/cm. In our silicon microring resonators with 5 μm radius, we caculate the loss of 5 by CST finite element simulation software is approximately 0.6 dB/cm. A silicon microring resonators with 1.5 μm radius has been fabricated in 2008 [46], thus, the dimensions of our silicon microring resonators can also be fabricated. The resonance equation of light is,

$$2\pi n_1 R - n_1 \times 50 + n_2 \times 50 = m\lambda \quad (m = 1, 2, 3...), \quad (11)$$

where, $n_1$ is the effective refractive index of the guided mode in the waveguide, $n_2$ is the effective refractive index of the guided mode in the slot, $m$ is the resonance order, and $\lambda$ is the wavelength of the resonant light. We calculate the ratio of the amplitude of the transmitted light and the incident light firstly. The calculation equations are shown in Eq. (12),

$$\begin{cases} E_2 = rE_1 + itE_3 \\ E_4 = rE_3 + itE_1 \\ E_6 = s_2 E_5 + s_1 E_{11} \\ E_7 = s_1 E_5 + s_2 E_{11} \end{cases} \begin{cases} E_5 = \tau e^{i\phi/2} E_4 \\ E_3 = \tau e^{i\phi/2} E_6 \\ E_8 = \tau e^{i\phi/2} E_7 \\ E_{11} = \tau e^{i\phi/2} E_{10} \end{cases}, \quad (12)$$

where $r$ and $t$ represent self-coupling coefficient and mutual coupling coefficient respectively, and satisfy $r^2 + t^2 = 1$. $s_1$ and $s_2$ represent reflectance and transmission, respectively.

Utilizing symbolic operations of Eq. (12), the ratio of the amplitude of the transmitted light to the incident light can be calculated by

$$\frac{E_2}{E_1} = \frac{r - (r^2 + 1)s_2 \tau^2 e^{i\phi} - r\tau^4 (s_1^2 - s_2^2) e^{2i\phi}}{1 - 2rs_2 \tau^2 e^{i\phi} - r^2 \tau^4 (s_1^2 - s_2^2) e^{2i\phi}}. \quad (13)$$

Similarly, the transmission of the exit port in microring resonator with a slot is calculated by

$$T = \left|\frac{r - (r^2 + 1)s_2 \tau^2 e^{i\phi} - r\tau^4 (s_1^2 - s_2^2) e^{2i\phi}}{1 - 2rs_2 \tau^2 e^{i\phi} - r^2 \tau^4 (s_1^2 - s_2^2) e^{2i\phi}}\right|^2. \quad (14)$$

## 3. RESULTS

### 3.1 The loss of waveguide varies with slot width under different sizes

Both in Fig. 1(a) and Fig. 1(d), the position of slot results in light scattering, leading to power losses. The light scattering includes diffuse and specular reflection and makes the light away from the resonator waveguide. When the power loss is significant at the slot, the light transmitted through the slot into the subsequent waveguide segment decreases. Additionally, a substantial amount of scattered light can induce higher-order modes. Adjusting the dimensions of the waveguide to reduce power losses at the slot is a prerequisite for investigating slotted micro-ring resonators. In this study, we focus on a microring resonator with a 50 nm slot opened at 12 clock on a 5 μm radius micro-ring. Since the circumference of the micro-ring, $\pi \times 10^4$ nm, is significantly larger than the slot size, we can treat the 50 nm slot at the 12 clock position in the micro-ring as a slot in a straight waveguide.

**Table 1 Structural parameters and refractive index of transverse slot waveguides**

| \ | Materials | Width (x-axis) | Height (y-axis) | Depth (z-axis) | Refractive index |
|---|---|---|---|---|---|
| Waveguide | Si | 700 nm | 450 nm | 2000 nm | 3.5 |
| Substrate | SiO$_2$ | 3000 nm | 6000 nm | 2000 nm | 1.44 |
| Cladding | Water | 3000 nm | 6000 nm | 2000 nm | 1.33 |
| Slot | Water | 700 nm | 450 nm | 50 nm | 1.33 |

Firstly, it is necessary to determine the waveguide dimensions of the straight waveguide and the micro-ring in the micro-ring resonator. The primary factor influencing power losses at the slot is the waveguide dimensions. Therefore, this study employs an arithmetic progression to establish the power losses for different waveguide dimensions. Specifically, the study investigates 6 structures with dimensions of 500×250 nm², 550×300 nm², 600×350 nm², 650×400 nm², 700×450 nm², and 725×475 nm², with an optical input wavelength of 1.55 μm and a slot width of sl. The study calculates the transmission (S21) and power losses (SS) for different transverse slotted waveguides, as shown in Fig. 2. The SS is the power loss in the microring resonator, and S21 is the power at the output port after normalization. Theoretically, the sum of the unnormalized S21 and the power loss in the straight waveguide and SS should be the input optical power.

Finite element simulation software is utilized to compute the transmission and power losses for the six different transverse slotted waveguide dimensions. As the slot width increases, the transmission gradually decreases while the power losses increase. Furthermore, it is observed that larger waveguide dimensions result in a slower reduction in transmission as the slot width increases, which is due to small divergence of the larger modal profile can help improve the coupling efficiency.

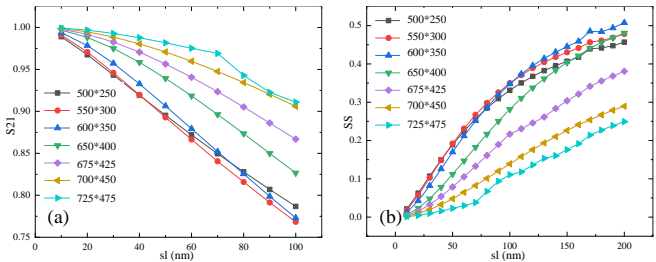

Fig. 2. Transmission (a) and power loss (b) in a transverse slot straight waveguide.

When the slot width is fixed at 50 nm, the power losses for the waveguide dimensions of 650×400 nm², 700×450 nm², and 725×475 nm² are -22.30 dB, -26.41 dB, and -32.84 dB, respectively. It is observed that for the waveguide dimension of 700×450 nm², the power loss is already below 0.05, indicating lower losses. However, larger waveguide dimensions make it more challenging to achieve efficient coupling between the straight waveguide and the microring, which is because the contrast between waveguide gaps achieving the same coupling coefficient is greater relative to large waveguide dimensions. Considering these two factors, this study decides to adopt the waveguide dimension of 700 × 450 nm² for the subsequent micro-ring resonator.

### 3.2 Optical field distribution with transverse slot

The parameters of the transverse slotted waveguide in Fig. 1(a) are shown in Table 1. Subsequently, based on the parameters in Table 1, simulation experiments are conducted. In this study, perfect matching layers are added around and at the back end of the structure to avoid the reflection of light back into the slot, which could affect the accuracy of the results.

The width of the slot along the $z$ direction is 50 nm, as shown in Fig. 1(a). Based on the aforementioned structure and parameters, the effective refractive index of the two fundamental modes within the waveguide are determined to be 2.364 and 1.959 for TM mode and TE mode, respectively. The effective refractive index of the two fundamental modes within the waveguide are determined to be 2.364 for TE mode and 1.959 for TM mode. Based on the TM and TE fundamental modes, the distributions of $E_x$, $E_y$, and $E_z$ at various positions are calculated, and the potential of the structure to confine the optical field is analyzed.

In this section, we first investigate the optical field distribution of the TM mode since the electric field along the $x$ direction, $E_x$, exhibits discontinuity. Therefore, in the TM mode, the main component of the electric field is the $E_x$ component. Fig. 3(a) and 3(b) show the $xy$ plane graph of the $E_x$ component and the $E_z$ component. From the graph, it is evident that the $E_x$ electric field undergoes a step change along the $x$ direction in the TM mode. Additionally, the electric field is weak at the upper and lower surfaces of the waveguide, while the electric field in the adjacent upper and lower claddings is relatively high. Since $E_z$ is the spatial derivative of $E_x$, it results in a significant $E_z$ electric field at the upper and lower boundaries of the waveguide.

Next, we analyze the $E_z$ component of the electric field in the slot, and the results are shown in Fig. 3(c-e). The results show that the $E_z$ component of the electric field in the slot can be clearly enhanced no matter in the $yz$ plane or the $xz$ plane. It is worth noting that in the $xy$ plane, the deepest blue and the deepest red are the places where the $E_z$ component of the electric field is the strongest, and at the upper and lower surfaces of the waveguide, the phase difference between the deepest blue and the deepest red is $\pi$.

In this section, we study the electric field distribution of the TE mode in the parametric structure of Table 1, because the electric field of the TE mode is discontinuous along the $y$ direction. In the TE mode, the main component of the electric field at this time is the $E_y$ component. Fig. 3(f) and 3(g) are the $xy$ plane diagrams of the $E_y$ component and the $E_z$ component. The results show that the $E_y$ of the TE mode is at the left and right boundaries of the waveguide along the $y$ direction. When a jump of electric field occurs, the electric field $E_z$ will be very large at the left and right boundaries of the waveguide. The results of Fig. 3(h-j) show that the $E_z$ component of the electric field in the slot can be clearly seen to be enhanced no matter in the *yz* plane or the *xz* plane.

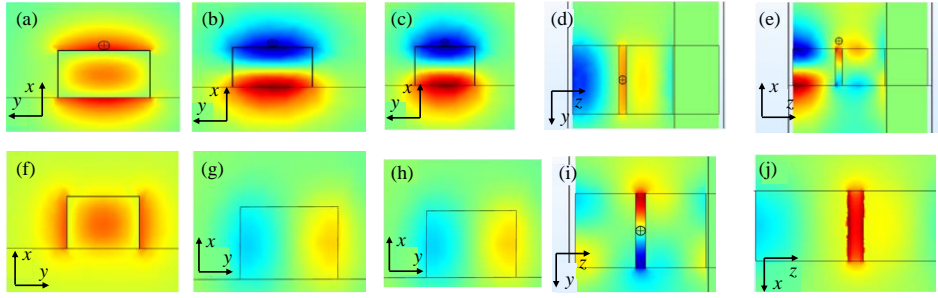

Fig.3. TM and TE fundamental mode electric field. TM fundamental mode electric field in the *xy* plane; (a): $E_x$; (b): $E_z$. Distribution of $E_z$ components in each plane under TM fundamental mode: *xy* plane (c), *yz* plane (d), *xz* plane (e). TE fundamental mode electric field in the *xy* plane; (f): $E_x$; (g): $E_z$. Distribution of $E_z$ components in each plane under TE fundamental mode: *xy* plane (h), *yz* plane (i), *xz* plane (j).

### 3.3 Optical force distribution with transverse slot

In this section, the distribution of optical force in TE and TM modes is analyzed. We consider a polystyrene nanoparticle (refractive index is 1.6) that is 50 nm in diameter. The nanoparticle that we consider here is nanoscale, random thermal motion of nanoparticles should not be ignored. We compare the distribution of force $F_x$ when the nanoparticle moves along the X direction (width) in TE and TM modes, as shown in Fig. 4.

The stable binding positions of TM mode and TE mode to the nanoparticle are at X = 445 nm and X = 210 nm respectively, which means that the TM mode binds the nanoparticle at the entrance of the slot, while the TE mode binds the nanoparticle to the slot entrance. The stable binding positions means where the nanoparticle in our novel structure can be trapped. At the stable binding position, the derivative of the optical force of the TM mode with respect to X is much larger than that of the TE mode, indicating that the optical field of the TM mode has a stronger binding ability to the particle nanoparticle.

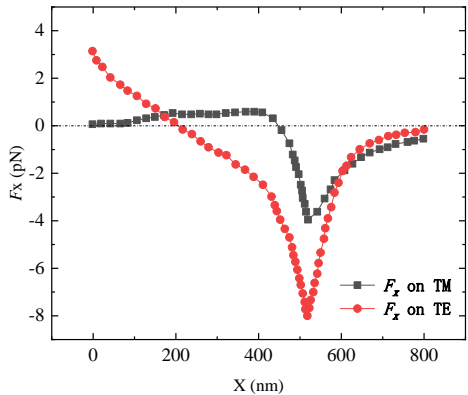

Fig.4. The $F_x$ of TM and TE modes *vs.* the X axis in transverse slot.

The maximum value of the optical force $F_x$ of the TE mode is about 3 times that of the TM mode. When the nanoparticle is above the waveguide and gradually away from the waveguide, the exponential decay of $F_x$ of the TE mode is faster than that of the TM mode, which indicates that the evanescent field decay of the TE mode is faster than that of the TM mode. This conclusion is consistent with the optical field distribution conclusion. Therefore, the TE fundamental mode is not conducive to the capture of particles by the waveguide structure. The calculations in this paper are based on the TM fundamental mode.

### 3.4 Resonant parameters and field distribution of microring resonator

In this section, we study the situation of the microring resonator without a slot firstly, and make the microring resonator reach the critical coupling state by adjusting the coupling distance between the straight waveguide and the microring in the finite element simulation software. The coupling distance between the straight waveguide and the microring at the critical coupling is 151 nm. The transmission and phase changes between 1.585 μm and 1.610 μm are calculated by parametrically scanning the wavelength, as shown in Fig. 5(a).

It can be viewed that when the incident wavelength is 1.597 μm, the transmission is 0.006, which can be considered to be approximately critical coupling. We calculate that the $Q^2$ is about 3013, and the FSR is 30.132 nm. Next, let the coupling distance between the straight waveguide and the microring be 151 nm, and the wavelength of the incident light be 1.597 μm. We calculate the power distribution and optical field distribution in this case.

In Fig. 6(a) and (b), this paper obtains the electric field modulus $|E|$ and the electric field Ey in the y direction at the time of critical coupling. It can be seen from Fig. 6(a) that the light is injected into the straight waveguide from the lower end, and then coupled into the ring to resonate in the ring, so there is an extinction phenomenon at the exit of the straight waveguide. It can be seen from Fig. 6(b) that the electric field $E_y$ is stronger at the 6 clock and 12 clock, and weaker at the 3 clock and 9 clock. This is the distribution characteristic of the electric field $E_y$. Similarly, it can be obtained that the electric field $E_z$ is weaker at 6 clock and 12 clock, and stronger at 3 clock and 9 clock. In Fig. 6(d), we can see the difference between the $E_y$ intensity of slotted microring resonator at 6 o'clock and 12 o'clock, because of the of the existence of slot.

### 3.5 Resonant parameters and field distribution of microring resonator with a slot

We design a 50 nm slot at 12 clock of the microring resonator, forming the structure as shown in Fig. 1(d). Let the coupling distance be 151 nm. Parametric sweep from 1.591 μm to 1.6 μm. The variation of transmission and phase with the wavelength of incident light is shown in Fig. 5(c) and (d). It can be seen from Fig. 5(c) that the transmission is 0.975 at the incident wavelength of 1.5951 µm, so the critical coupling is not reached at the coupling distance of 151 nm, and the phase in Fig. 5(d) does not have a π phase abrupt change at 1.595 µm. Therefore, the microring resonator is under-coupled at this time[37]. At this time, the $Q^2$ value of the resonator is 13293, and the FSR is 30.089 nm.

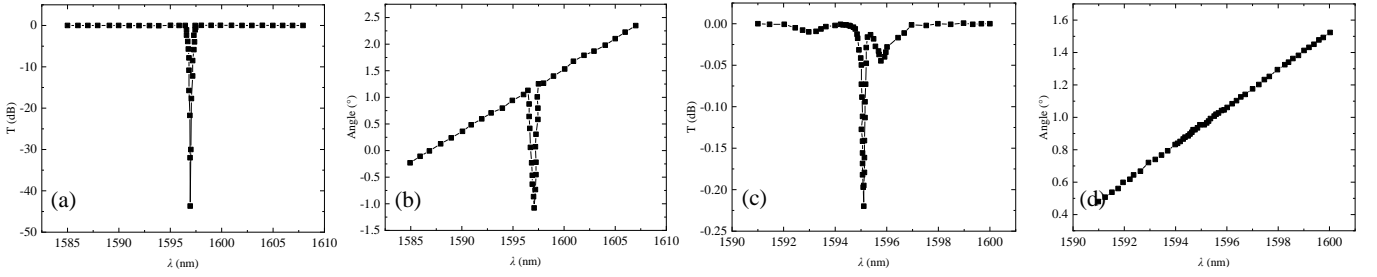

Fig.5. Transmission and phase *vs.* wavelength. Transmission (a) and phase (b) of microring resonator without slot. Transmission (c) and phase (d) of microring resonator with a slot.

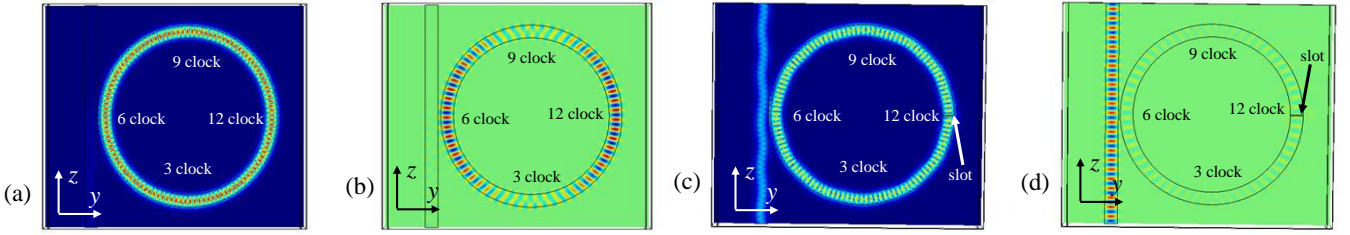

Fig.6. The distribution of the electric field norm $|E|$ and the electric field $E_y$. Electric field mode $|E|$(a) and electric field $E_y$(b) at critical coupling in the microring resonator. The electric field mode $|E|$(c) and the electric field $E_y$(d) of the microring resonator with a slot.

Then, we reduce the coupling distance to achieve critical coupling. Firstly, the coupling distance is fixed at 151 nm and the incident light wavelength is 1.5951 µm. The optical field distribution at this time is calculated as shown in Fig. 6(c) and (d). It can be seen from Figure 6(c) that a small part of the incident light enters the ring to resonate, and most of the light does not enter the microring. The electric field distribution in Fig. 6(d) verifies this again.

### 3.6 Calculation of optical force in microring resonator with a 50 nm slot

In this section, we calculate the bound force $F_x$ in the X direction of the nanoparticle with a diameter of 50 nm and a refractive index of 1.6 in the microring resonator [47]. The bound force of parametrically scanned nanoparticle along the width of the waveguide is shown in Fig. 7. At 12 clock of microring resonator, when Y = 0 nm, the nanospheres are on the upper surface of microring waveguide and are located in the exact center of the slot. In Fig. 7, the black curve is the bound force $F_x$ of the nanoparticle in the X direction on the straight waveguide with a slot, and the blue and red curves are the resonant force of nanoparticle in microring resonator with and without slot, respectively. The bound force in the X direction on the structure.

It can be viewed from Fig. 7 that the maximum optical force experienced by the nanoparticle on the slot (Y = -260 nm) is 1.74 pN/W, and the nanoparticle in microring resonator (Y = 60 nm, S21 = 0.6203). The maximum optical force on microring waveguide with the slot (Y = 20 nm, S21 = 0.6115) is 3988.8 pN/W, which is 2292 times on the slot straight waveguide. The optical force of our novel structure is 2.266 times of microring resonator without slot, which is the typical value of this bound optical force. This structure greatly improves the ability of the beam to bind nanoparticles.

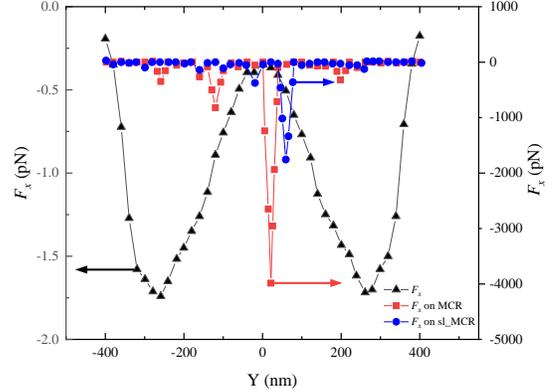

Fig.7. The distribution of optical force when the nanoparticle with and without slot moves along the Y direction on the waveguide.

## 4. DISCUSSION

Due to the limitation of simulation time and computer processing memory, the radius of the microring simulated by finite element simulation software in this paper is only 5 µm, and the bending loss of this microring will be relatively large. In the actual manufacturing process, the radius of the microring will be larger, so the bending loss will be further reduced, and the binding effect on the nanoparticles will be enhanced.

For the dispersion problem, because the effective refractive index of the guided mode in the microring will change with the wavelength of the incident light, and the effective resonance radius of the light in the microring also change with the wavelength of the incident light and the added 50 nm slot [48]. The scattering in the microring is dominated by nonlinear dispersion. Since this paper only uses numerical calculation software for simple verification, this paper does not consider the dispersion problem in the numerical calculation software simulation at present, and these factors are all in the finite element simulation software taken into account automatically.

The results of the numerical calculation software may be slightly different from the results of the finite element simulation software. In the later stage, we can focus on studying the dispersion in the microring resonator, and bring the dispersion equation into the simulation program of the numerical calculation software to improve the accuracy of the numerical calculation software results.

To evaluate the stability of the efficient trapping, the increase of temperature should be taken into account [44]. However, we can ensure that the optical trapping is a constant temperature in the numerical simulation and the next step of the experiment. Our constant temperature chamber can make the trapping process within a temperature change of less than 0.1 °C. In this case, we will ensure that the temperature of the trapping process is 10 °C, and no longer numerically simulate the influence of temperature changes on optical trapping.

## 5. CONCLUSION

In summary, our study focuses on a structure that incorporates a 50 nm slot positioned at the 12 clock position of the microring resonator. Through extensive analysis, we have demonstrated that this structure exhibits significant field enhancement at the slot region. In this paper, we have employed finite element simulation software and numerical calculation methods to investigate the spectral characteristics of both the microring resonator and the microring structure with the slot. Our findings reveal that the structure achieves a substantial increase in the optical force exerted on the slot. Specifically, for a nanosphere with a refractive index of 1.6 and a diameter of 50 nm, we have observed a maximum optical force on the slot of 3988.8 pN/W. This value is 2292 times higher than the maximum optical force observed on the straight waveguide. Furthermore, it is 2.266 times higher than the maximum optical force received in the medium. The significant enhancement in the binding ability of nanoparticles provided by this structure opens up new possibilities for more sophisticated micro-nano manipulation techniques. This research paves the way for the development of advanced technologies in the field of micro-nano manipulation.

**Funding.** This research received no external funding.

**Acknowledgments.** This work was supported by School of Precision Instruments and Optoelectronic Engineering, Tianjin University and Beijing Tsingmicro Intelligent Technology Co. Ltd. We acknowledge support by Tianjin University and Beijing Tsingmicro Intelligent Technology Co. Ltd.

**Disclosures**. The authors declare no conflicts of interest.

**Data availability**. Data underlying the results presented in this paper are not publicly available at this time but maybe obtained from the authors upon reasonable request.

**Supplemental document**. There is no supplemental file for this paper.

## References

1. K. Svoboda, and S. M. Block, "Biological applications of optical forces," Annu. Rev. Biophys. Biomolec. Struct., **23**, 247-285 (1994).
2. A. Ashkin, "Acceleration and trapping of particles by radiation pressure," Phys. Rev. Lett., **24**, 156-159 (1970).
3. X. Zhou, O. Wu, "Implicit counterfactual data augmentation for deep neural networks," arXiv 2023, arXiv: 2304.13431.
4. R. Omori, T. Kobayashi, and A. Suzuki, "Observation of a single-beam gradient-force optical trap for dielectric particles in air," Opt. Lett., **22**, 816-818 (1997).
5. D.G. Grier, "A revolution in optical manipulation," Nature, **424**, 810-816 (2003).
6. B. S. Schmidt, A. H. J. Yang, D. Erickson, and M. Lipson. "Optofluidic trapping and transport on solid core waveguides within a microfluidic device," Opt. Express, **15**, 14322-14334 (2007).
7. S. Mandal, X. Serey, and D. Erickson. "Nanomanipulation using near field photonics," Lab Chip, **11**, 995-1009 (2011).
8. L. N. Ng, M. N. Zervas, J. S. Wilkinson, and B. J. Luff. "Manipulation of colloidal gold nanoparticles in the evanescent field of a channel waveguide," Appl. Phys. Lett., **76**, 1993-1995 (2000).
9. S. Kawata, and T. Tani, "Optically driven Mie particles in an evanescent field along a channeled waveguide," Opt. Lett., **21**, 1768-1770 (1996).
10. X. Zhou, N. Yang, O. Wu, "Combining adversaries with anti-adversaries in training," In Proceedings of the AAAI Conference on Artificial Intelligence, 2023.
11. K. Grujic, O. G. Helleso, J. P. Hole, and J. S. Wilkinson, "Sorting of polystyrene microspheres using a Y-branched optical waveguide," Opt. Express, **13**, 1-7 (2005).
12. X. Zhou, W. Ye, Y. Wang, C. Jiang, Z. Lee, R. Xie, S. Zhang, "Enhancing in-context learning via implicit demonstration augmentation," arXiv 2024, arXiv: 2407.00100.
13. V. R. Almeida, Q. Xu, C. A. Barrios, and M. Lipson. "Guiding and confining light in void nanostructure," Opt. Lett., **29**, 1209-1211 (2004).
14. X. Zhou, O. Wu, W. Zhu, Z. Liang, "Understanding difficulty-based sample weighting with a universal difficulty measure," In Proceedings of the Joint European Conference on Machine Learning and Knowledge Discovery in Databases, 2022.
15. A. H. J. Yang, S. D. Moore, B. S. Schmidt, M. Klug, M. Lipson, and D. Erickson. "Optical manipulation of nanoparticles and biomolecules in sub-wavelength slot waveguides," Nature, **457**, 71-75 (2009).
16. J. B. Driscoll, X. Liu, S. Yasseri, I. Hsieh, J. I. Dadap, and R. M. O. Jr. "Large longitudinal electric fields ($E_z$) in silicon nanowire waveguides," Opt. Express, **17**, 2797-2804 (2009).
17. M. Hoshina, N. Yokoshi, H. Okamoto, and H. Ishihara. "Super-resolution trapping: A nanoparticle manipulation using nonlinear optical response," ACS Photonics, **5**, 318-323 (2018).
18. H. Minamimoto, N. Oyamada, and K. Murakoshi, "Toward room-temperature optical manipulation of small molecules," J. Photochem. Photobiol. C-Photochem. Rev., **24**, 100582 (2023).
19. C. Pin, H. Fujiwara, and K. Sasaki, "Controlled optical manipulation and sorting of nanomaterials enabled by photonic and plasmonic nanodevices," J. Photochem. Photobiol. C-Photochem. Rev., **52**, 100534 (2022).
20. T. W. Ebbesen, H. J. Lezec, H. F. Ghaemi, T. Thio, and P. A. Wolff. "Extraordinary optical transmission through sub-wavelength hole arrays," Nature, **391**, 667-669 (1998).
21. L. Hao, X. Yu, X. Wu, W. Shi, M. Chen, L. Liu, and L. Xu. "All-optically-controlled nanoparticle transporting and manipulating at SOI waveguide intersections," Opt. Express, **20**, 24160-24166 (2012).
22. Y. Shi, H. Zhao, L. K. Chin, Y. Zhang, P. H. Yap, W. Ser, C.-W. Qiu, and A. Q. Liu. "Optical potential-well array for high-selectivity, massive trapping and sorting at nanoscale," Nano Lett., **20**, 5193-5200 (2020).
23. L. Fang, and J. Wang, "Optical trapping separation of chiral nanoparticles by subwavelength slot waveguides. Phys. Rev. Lett., **127**, 233902 (2021).
24. X. Zhou, O. Wu, M. Li, "Investigating the sample weighting mechanism using an interpretable weighting framework," IEEE Trans. Knowl. Data Eng., **36**, 2041-2055 (2023).
25. L. Novotny, R. X. Bian, and X. S. Xie, "Theory of nanometric optical tweezers," Phys. Rev. Lett., **79**, 645-648 (1997).

26. D. G. Kotsifaki, M. Kandyla, and P. G. Lagoudakis, "Near-field enhanced optical tweezers utilizing femtosecond-laser nanostructured substrates," Appl. Phys. Lett., **107**, 211111 (2015).
27. R.R. Gutha, S. M. Sadeghi, C. Sharp, and W. J. Wing. "Biological sensing using hybridization phase of plasmonic resonances with photonic lattice modes in arrays of gold nanoantennas," Nanotechnology, **28**, 355504 (2017).
28. A. Morita, T. Sumitomo, A. Uesugi, K. Sugano, and Y. Isono. "Dynamic electrical measurement of biomolecule behavior via plasmonically-excited nanogap fabricated by electromigration," Nano Express, **2**, 010032 (2021).
29. J. Li, Y. Ma, Y. Gu, I.-C. Khoo, and Q. Gong. "Large spectral tunability of narrow geometric resonances of periodic arrays of metallic nanoparticles in a nematic liquid crystal," Appl. Phys. Lett., **98**, 213101 (2011).
30. D. Yoo, K. L. Gurunatha, H.-K. Choi, D. A. Mohr, C. T. Ertsgaard, R. Gordon, and S.-H. Oh. "Low-power optical trapping of nanoparticles and proteins with resonant coaxial nanoaperture using 10 nm gap," Nano Lett., **18**, 3637-3642 (2018).
31. V. G. Kravets, A. V. Kabashin, W. L. Barnes, and A. N. Grigorenko. "Plasmonic surface lattice resonances: A review of properties and applications," Chem. Rev., **118**, 5912-5951 (2018).
32. G. Brunetti, N. Sasanelli, M. N. Armenise, and C. Ciminelli. "Optically driven Mie particles in an evanescent field along a channeled waveguide," Photonics, **8**, 425 (2022).
33. Y. Pang, and R. Gordon. "Optical trapping of a single protein," Nano. Lett., **11**, 402-406 (2012).
34. H. Li, W. Song, Y. Zhao, Q. Cao, and A. Wen. "Optical trapping, sensing, and imaging by photonic nanojets," Photonics, **8**, 434 (2021).
35. A. H. J. Yang, and D. Erickson, "Optofluidic ring resonator switch for optical particle transport," Lab Chip, **10**, 769-774 (2010).
36. Y. Geng, T. Zhu, H. Lv, Y. Cao, F. Sun, and W. Ding. "Flexible optical manipulation of ring resonator by frequency detuning and double-port excitation," Opt. Express, **24**, 15863-15871 (2016).
37. D. G. Kotsifaki, V. G. Truong, and S.N. Chormaic, "Fano-resonant, asymmetric, metamaterial-assisted tweezers for single nanoparticle trapping," Nano Lett., **20**, 3388-3395 (2020).
38. X. Zhou, H. Tamura, T.-H. Chang, and C.-L. Hung. "Coupling single atoms to a nanophotonic whispering-gallery-mode resonator via optical guiding," Phys. Rev. Lett., **130**, 103601 (2023).
39. L. Zhou, K. Okamoto, and S. J. B. Yoo. "Athermalizing and trimming of slotted silicon microring resonators with UV-sensitive PMMA upper-cladding," Photon. Technol. Lett., **21**, 1175-1177 (2009).
40. L. Fang, and J. Wang. "Optical trapping separation of chiral nanoparticles by subwavelength slot waveguides," Phys. Rev. Lett., **127**, 233902 (2021).
41. D. Conteduca, G. Brunetti, G. Pitruzzello, F. Tragni, K. Dholakia, T. F. Krauss, and C. Ciminelli. "Exploring the limit of multiplexed near-field optical trapping," ACS photonics, **8**, 2060-2066 (2021).
42. D. G. Kotsifaki, V. G. Truong, and S. N. Chormaic. "Fano-resonant, asymmetric, metamaterial-assisted tweezers for single nanoparticle trapping," Nano Lett., **20**, 3388-3395 (2020).
43. S. Lin, and K. B. Crozier, "Planar silicon microrings as wavelength-multiplexed optical traps for storing and sensing particles," Lab Chip, **11**, 4047-4059 (2011).
44. X. Zhou, O. Wu, N. Yang, "Adversarial Training With Anti-Adversaries", IEEE Trans. Pattern Anal. Mach. Intell., **46**, 10210-10227 (2024).
45. J. E. Heebner, V. Wong, A. Schweinsberg, R. W. Boyd, and D. J. Jackson. "Optical transmission characteristics of fiber ring resonators," IEEE J. Quantum Electron., **40**, 726-730 (2004).
46. Q. Xu, D. Fattal, and R. Beausoleil. "Silicon microring resonators with 1.5-µm radius," Opt. Express, **16**, 4309-4315 (2008).
47. S. Lin, E. Schonbrun, and K. Crozier, "Optical manipulation with planar Silicon microring resonators," Nano Lett., **10**, 2408-2411 (2010).
48. V. E. Lobanov, A. V. Cherenkov, A. E. Shitikov, I. A. Bilenko, and M. L. Gorodetsky. "Dynamics of platicons due to third-order dispersion," Eur. Phys. J. D, **71**, 185-193 (2017).